\documentclass{article}
\usepackage{amsmath,epsfig}
\usepackage[preprint]{spconfa4}
\usepackage{xcolor}
\usepackage{graphicx}
\usepackage{url}
\usepackage{amssymb}

\let\OLDthebibliography\thebibliography
\renewcommand\thebibliography[1]{
  \OLDthebibliography{#1}
  \setlength{\parskip}{0pt}
  \setlength{\itemsep}{0pt plus 0.3ex}
}

\begin{document}\sloppy

\def\x{{\mathbf x}}
\def\L{{\cal L}}

\title{Object-Based Image Coding: A Learning-Driven Revisit}
%
\name{Qi Xia, Haojie Liu, and Zhan Ma\thanks{Z. Ma is the corresponding author.}}
\address{Vision Lab, Nanjing University\\
\{qi\_xia, haojie\}@smail.nju.edu.cn, mazhan@nju.edu.cn}

\maketitle

\begin{abstract}
{The Object-Based Image Coding (OBIC) that was extensively studied about two decades ago, promised a vast application perspective for both ultra-low bitrate communication and high-level semantical content understanding, but it had rarely been used due to the inefficient compact representation of object with arbitrary shape. A fundamental issue behind is how to efficiently process the arbitrary-shaped objects at a fine granularity (e.g., feature element or pixel wise). To attack this, we have proposed to apply the element-wise {\it masking} and {\it compression} by devising an object segmentation network for image layer decomposition, and parallel  convolution-based neural image compression networks to process masked foreground objects and background scene separately. All  components are optimized in an end-to-end learning framework to intelligently weigh their (e.g., object and background) contributions for visually pleasant reconstruction. We have conducted comprehensive experiments to evaluate the performance on PASCAL VOC  dataset at a very low bitrate scenario (e.g., $\lesssim$ 0.1 bits per pixel - bpp) which have demonstrated noticeable subjective quality improvement compared with JPEG2K, HEVC-based BPG and another learned image compression method. All relevant materials are made publicly accessible at \url{https://njuvision.github.io/Neural-Object-Coding/}.}

\end{abstract}
\section{Introduction}
\label{sec:intro}
Visual information accounts for more than 70\% of all sensory input in our human body\footnote{\url{https://antranik.org/the-eye-and-vision/}}. A typical representative example of visual information illustration is an {\it image} where  objects (e.g., face, car plate, building, etc), background, and their relative layered associations are presented to convey information. Ideally, image is a composite model of relevant objects by which appropriate visual perception (e.g., depth, geometry, photometric attributes, etc) and content understanding (e.g., segmentation, classification, etc) based applications are performed.

Therefore, a straightforward application-driven image restoration is to reconstruct its embedded and meaningful objects, rather the entire scene, which has promised a great potential for both low bitrate and high-level semantical applications. It had  led to extensive explorations on object-based image coding (OBIC) about two decades ago. However, it had rarely been applied in practice, even with a industry standard (e.g., MPEG-4~\cite{puri1998mpeg}) concluded at that time. This is mainly due to the inefficient compact object representation and insufficient computing capability back to early 2000's. For example, although shape-adaptive discrete wavelet and  cosine transforms~\cite{sikora1995shape,li2000shape} were dedicatedly developed for  arbitrary  shape support, they introduced a great amount of computational cost for implementation.

Recently, thanks to the advances in algorithm and hardware of deep learning~\cite{lecun2015deep}, we have envisioned the emerging breakthrough of such classical OBIC problem. Here, key issue behind is how to efficiently segment and compress objects from an image, at the element or pixel granularity to support arbitrary-shaped representation. Thus, we have proposed to apply the element-wise {\it masking} mechanism to activate or deactivate pixels for arbitrary-shaped object processing which can be easily coupled with regular stacked convolutions to efficiently exploit local spatial correlation, without resorting for dedicated shape-adaptive tool design as in~\cite{sikora1995shape, li2000shape}.

\begin{figure*}[t]
	\centering
	\includegraphics[scale=0.75]{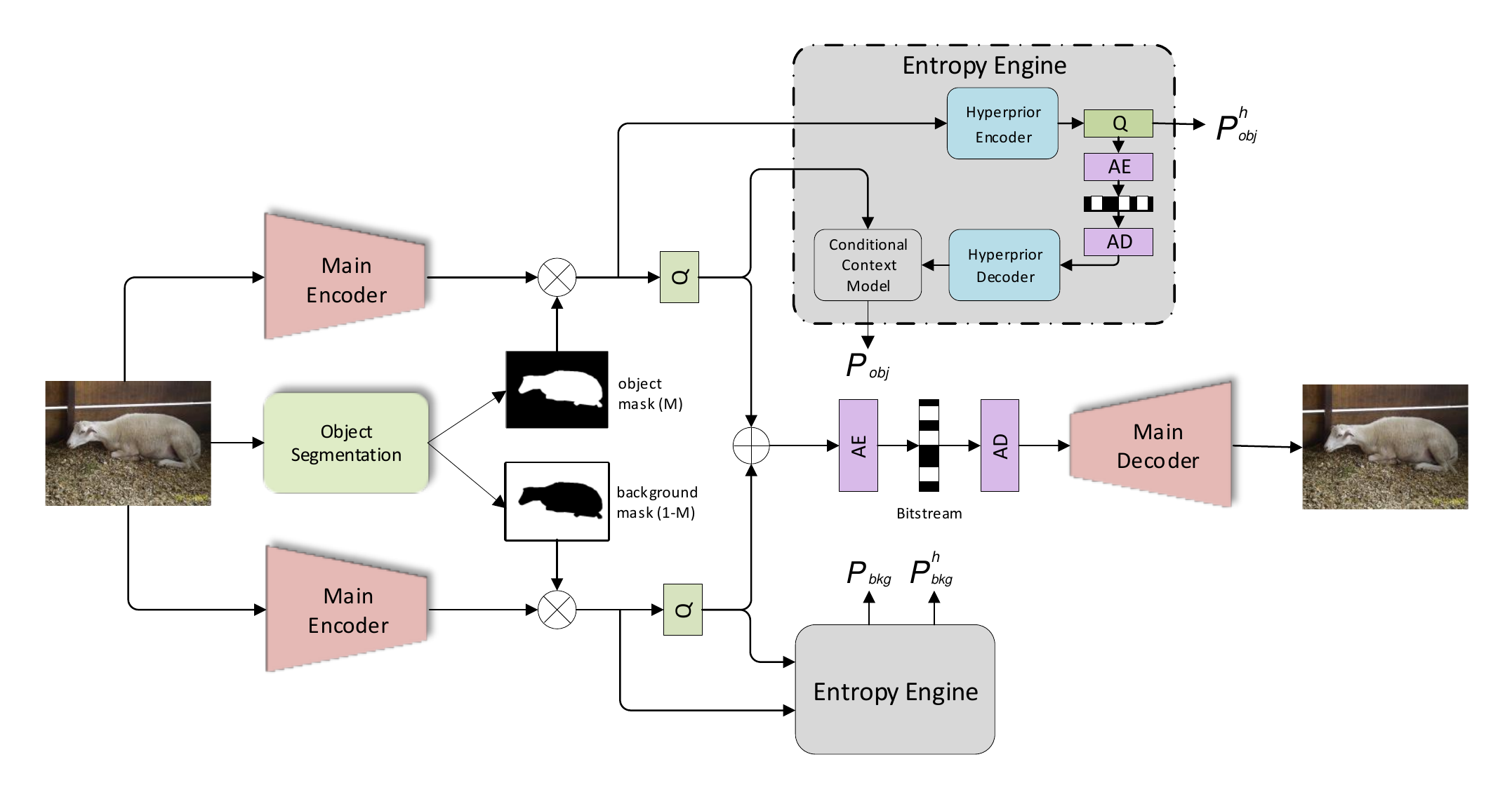}
	\caption{{\bf LearntOBIC.} A two-layer structure is exemplified with upper part for {\it object} and  lower part for {\it background}. Image segmentation uses the DeepLab~\cite{chen2017deeplab} with ResNet-34~\cite{he2016deep} as backbone, while a variational autoencoder (VAE) based NLAIC~\cite{chen2019neural} is used to process masked image layers in parallel.  Q is for quantization, AE and AD are for respective arithmetic encoding and decoding. Entropy engine shares the identical structure for object and background layer where factorized context probability is used for hyper feature maps (fMaps) ${\bf F}^h$, and context-adaptive probability estimation is  for latent fMaps $\bf F$. }
	\label{fig:framework}
\end{figure*}

{Towards this goal, we choose to use the segmentation network (e.g., DeepLab) in~\cite{chen2017deeplab} to derive the element-wise mask which is utilized to produce masked image layers (e.g., object, or background) for subsequent parallel image compression and reconstruction\footnote{For simplicity, we have devised a binary mask, for a foreground object layer and a background layer. Multiple image layers can be supported easily.} {using stacked convolutional neural network (CNN) based NLAIC in~\cite{chen2019neural} }that can efficiently leverage the properties of masking and convolutional operations, for high-efficiency shape-adaptive object processing.}
All components are connected and optimized in an end-to-end learning framework, as shown in Fig.~\ref{fig:framework}, where masked image layers are processed in parallel.  This system is referred to as the {\bf LearntOBIC}. Rate-distortion optimization (RDO) is conducted to maximize the visual quality at a given bitrate budget.  

Our LearntOBIC is then evaluated using the public PASCAL VOC 2012 and Kodak dataset for low bitrate application scenarios, in comparison to the existing JPEG 2000~\cite{skodras2001jpeg}, High-Efficiency Video Coding (HEVC)-based Image Compression (aka, BPG)~\cite{sullivan2012overview} and NLAIC~\cite{chen2019neural}, offering significant visual quality improvement for  object reconstruction. Ablation studies are also offered to further discuss the capacity of proposed system.

To the best of our knowledge, this is the {\it first} work to revisit the OBIC via an end-to-end learning approach. Key novelties of this paper include: 1) {\bf End-to-end learnable} framework for intelligent object-based compression by parallel processing masked image layers; 2) {\bf Element-wise masking} mechanism to support arbitrary-shaped object  processing that  leverages the advances in object segmentation and CNN-based image compression; 3) {\bf Modularized functional} components for supporting parallel processing and future extension (e.g., multiple image layers, different utility loss, sub-stream extraction, etc).

\pagenumbering{gobble}
\section{Related Work} \label{sec:related_work}
We have categorized relevant research activities into three major classes, i.e., image/object segmentation, learnt image compression and non-uniform image encoding, mainly with the focus on recent learning-driven approaches.

{\bf Image/Object Segmentation.}
 Image semantic segmentation is aimed to classify image pixels into different (object) instances. Long {\it et al.}~\cite{Long_2015_CVPR} was the first one to introduce an end-to-end fully convolutional network (FCN) for image segmentation, which was then improved by follow-up works, such as the U-Net~\cite{ronneberger2015u}, feature pyramid network (FPN)~\cite{lin2017feature}, DeepLab~\cite{chen2017deeplab}, etc. DeepLab is used in this work  to produce accurate element-wise object segmentation mask, due to its superior performance introduced by the atrous convolution-based field-of-view enlargement, spatial pyramid pooling-based multiscale image context combination, and fully connected Conditional Random Fields enhanced localization of object boundaries.
 In the subsequent experimental studies, we utilize the ResNet-34 as the backbone for object  mask generation.

{\bf Learnt Image Compression.}
Deep CNN-based image compression methods have been studied extensively and shown {a promising and encouraging  progress on coding efficiency.} Such compression efficiency improvements are mainly contributed by the VAE with autoregressive neighbors and hyperprior~\cite{minnen2018joint,chen2019neural}, nonlinear transform (e.g., generalized divisive normalization - GDN~\cite{balle2016end}, non-local operations~\cite{chen2019neural}), differentiable quantization~\cite{balle2018variational}, embedded attention mechanisms~\cite{li2018learning, agustsson2019generative,chen2019neural}, etc. Recently, {learnt image compression methods even have emerged with better  performance than HEVC-based BPG, }with image quality measured by both PSNR and MS-SSIM~\cite{minnen2018joint,chen2019neural}.


{\bf Object-Based Non-uniform Image Encoding.}
Images are used to convey visual information. Either human or machine (e.g., autonomous drive) cares about the embedded objects with high-quality reconstruction for better perception and understanding. Thus,
non-uniform image encoding has been investigated to allocate different bitrate budgets for objects or region-of-interests ~\cite{han, han2008object, chen2006dynamic}  with non-uniform reconstructed qualities. However, performance always suffered because it was difficult to implement pixel-wise object representation in such conventional block-based system, and to apply the joint optimization between segmentation and compression.

Recently, attention mechanism offers the flexibility to apply fine-grained weights especially at object edges and rich textures using {\it masks}  when integrated with the learnt image compression methods~\cite{li2018learning, agustsson2019generative, chen2019neural}, leading to the noticeable compression efficiency improvement.

\begin{figure*}
	\centering
	\includegraphics[scale=0.7]{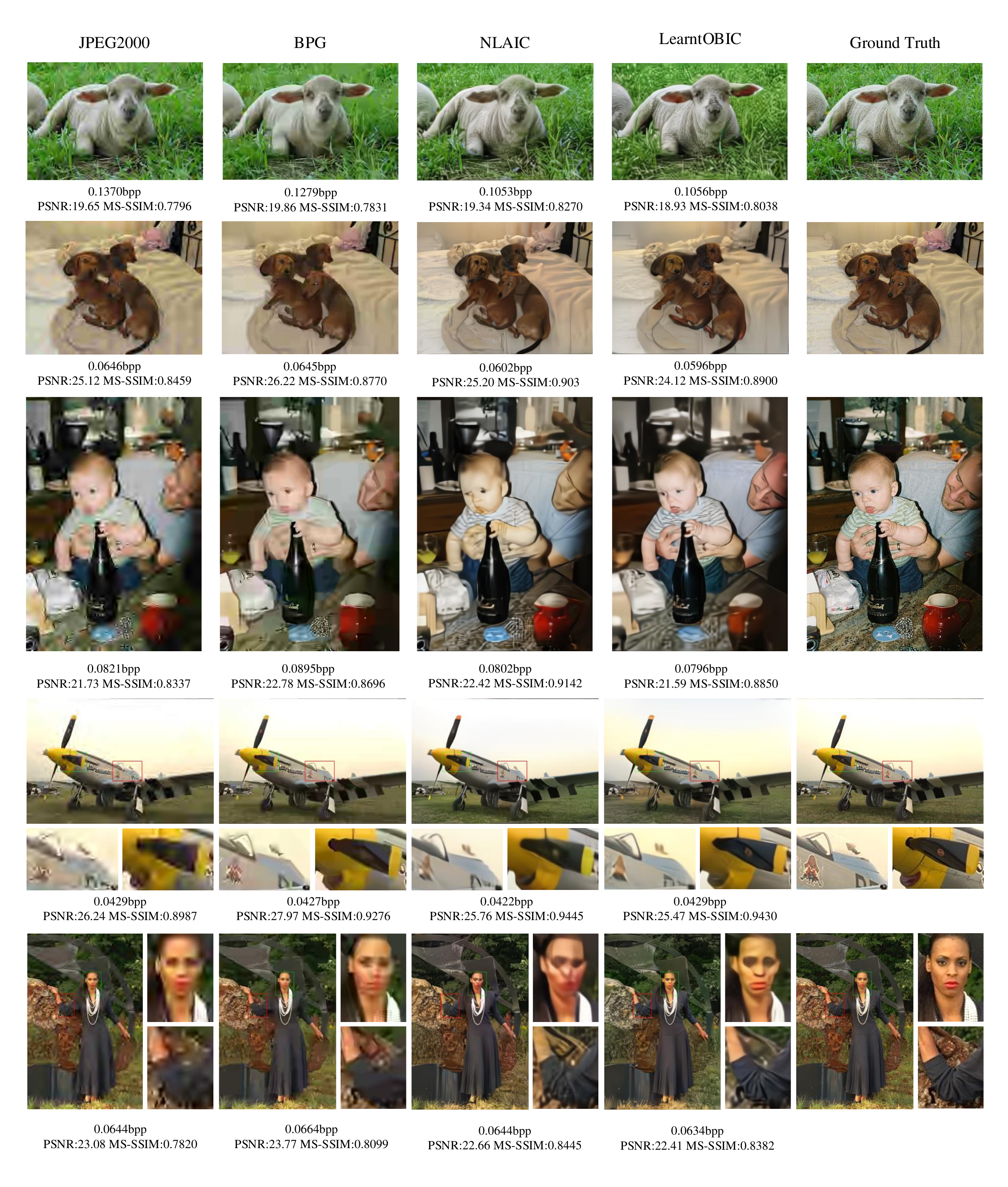}
	\caption{{\bf Visual Comparison} Reconstructed snapshots processed by JPEG2K, BPG, NLAIC~\cite{chen2019neural} and proposed LearntOBIC using PASCAL VOC 2012 and Kodak test images. Our method achieves significantly better subjective quality. PSNR, SSIM and total bit rate are also marked. {\bf Zoom in the pictures, and you will see more image details.}}
	\label{fig:visual_result}
\end{figure*}

\section{LearntOBIC: Object-Based Image Compression via End-to-End Learning} \label{sec:learnt_OBIC}
Leveraging the advances in learning-based image segmentation, compression and quality control (as discussed in Section~\ref{sec:related_work}), we have proposed to integrate the segmentation network with image compression network in a fully end-to-end learnable framework, for efficient object-based image coding. We have exemplified our LearntOBIC using a two-layer decomposition structure including the (foreground) object and background, as shown in Fig.~\ref{fig:framework}.

\subsection{Object Mask Generation}
We use DeepLab~\cite{chen2017deeplab} to classify input image for the element-wise  mask  derivation. As revealed in our simulations, object segmentation accuracy in compression task is not as crucial as in other vision tasks (e.g., recognition). Thus, we choose to use ResNet-34 as the backbone for DeepLab to generate masks, instead of the original ResNet-101, considering the balanced tradeoff between the accuracy and computational complexity. As shown in Table~\ref{tab:mIOU}, there is negligible compression efficiency loss for applying the ResNet-34 with $\approx 10\%$ segmentation accuracy drop measured using mIOU, but  2$\times$ speedup with half GFLOPs requirement, in comparison to the RestNet-101. Our LearntOBIC can be easily extended to support masks generated by various segmentation works.

 \begin{table}[t]
  \centering
  \caption{Segmentation Accuracy (mIOU $^\star$) and Complexity (GFLOPs) of ResNet-34 and ResNet-101 in DeepLab~\cite{chen2017deeplab}}
  \begin{tabular}{|c|c|c|}
    \hline
        & mIOU & GFLOPs \\
    \hline
     ResNet-34 & 68.00\% &  135.94 \\
     \hline
     ResNet-101 & 77.69\% & 267.57 \\
    \hline
  \end{tabular}\\
  $^\star$Mean Intersection over Union - mIOU
  \label{tab:mIOU}
\end{table}

For the two-layer structure, we are using the binary mask to indicate the feature elements (e.g., pixels) corresponding to the foreground object and background, respectively. We refer $\bf M$ as the object mask while (1-$\bf M$) is the background mask. Masks can be augmented with image in its pixel domain or feature domain, via element-wise multiplication. Here, we choose to apply the mask in feature domain to derive corresponding activated fMaps of respective object and background,
\begin{align}
    {\bf F}_{\tt obj} & = {\bf M} \otimes {\bf F}, \mbox{~~~}
    {\bf F}_{\tt bkg} = (1 - {\bf M}) \otimes {\bf F}.
\end{align} $\bf F$ is the latent fMaps generated by NLAIC encoder in Fig.~\ref{fig:framework} and $\otimes$ is the element-wise multiplication. Note that $\bf M$ presents the same height $H$ and width $W$ as $\bf F$ at a size of $H\times W\times C$ with $C$ for the number of channels. Identical mask $\bf M$ or ($1- {\bf M}$) is applied to all channels. Masked fMaps will then be processed in parallel for independent quantization and context modeling using hyper encoder-decoder pairs. 

\subsection{Parallel Image Layer Compression}
We use NLAIC as 
the basic codec unit to process the object and background layers. NLAIC uses the popular stacked CNN-based VAE structure with both hyperpriors and autoregressive neighbors for context modeling. 

Masked fMaps, i.e., ${\bf F}_{\tt obj}$ and ${\bf F}_{\tt bkg} $, are fed into hyper encoder-decoder pairs for accurate context modeling.  Conditional probability ${\bf P}_{\tt obj}$ or ${\bf P}_{\tt bkg}$ for latent fMaps, and factorized probability ${\bf P}_{\tt obj}^h$ or ${\bf P}_{\tt bkg}^h$ for hyper fMaps,  are applied to respective object and background layer for accurate entropy rate estimation that will be devised for RDO and actual binary bits  generation.

Rate estimation is used for object or background layer individually, each of which will have rate dissipated at both latent and hyper fMaps, i.e., 
\begin{align}
  R_{\tt obj} = -\sum \log_2({\bf P}_{\tt obj}) - \sum\log_2({\bf P}_{\tt obj}^h),\label{eq:obj_rate}\\
  R_{\tt bkg} = -\sum \log_2({\bf P}_{\tt bkg}) - \sum\log_2({\bf P}_{\tt bkg}^h), \label{eq:bkg_rate}
\end{align} with $\sum$ indicating the entropy sum by traversing all feature elements.
On the contrary, we will use the total distortion between compressed result and input image, e.g., $D_{\tt tot} = D({I}_{\tt out},  {I}_{\tt in})$,  that can be measured using PSNR, MS-SSIM or even feature loss for end-to-end learning.

For implementation, we can put compressed object and background layers in separated sub-streams for subsequent multiplexing, by which we can offer the individual layer reconstruction by extracting specific sub-stream. This would generally enable the capability for object-based tasks, without streaming and decoding the entire image.

\begin{table}[t]
  \centering
  \caption{Objective result on PASCAL VOC 2012 test set.}
  \begin{tabular}{|c|c|c|c|c|}
    \hline
        & JPEG2K & BPG & NLAIC & LearntOBIC \\
    \hline
    ave. bpp & 0.0675 & 0.0671 & 0.0636 &  0.0648 \\
     \hline
     MS-SSIM & 0.8339 & 0.8669 & \color{red}{\bf 0.9083} & \color{blue}{\bf 0.8992} \\
    \hline
     PSNR & 23.27 & {\color{red}{\bf 24.08}} & \color{blue}{\bf 23.36} & 22.71\\
    \hline
  \end{tabular}
  \label{tab:BDrate}
\end{table}

\begin{figure}[b]
\centering
	\includegraphics[scale=0.5]{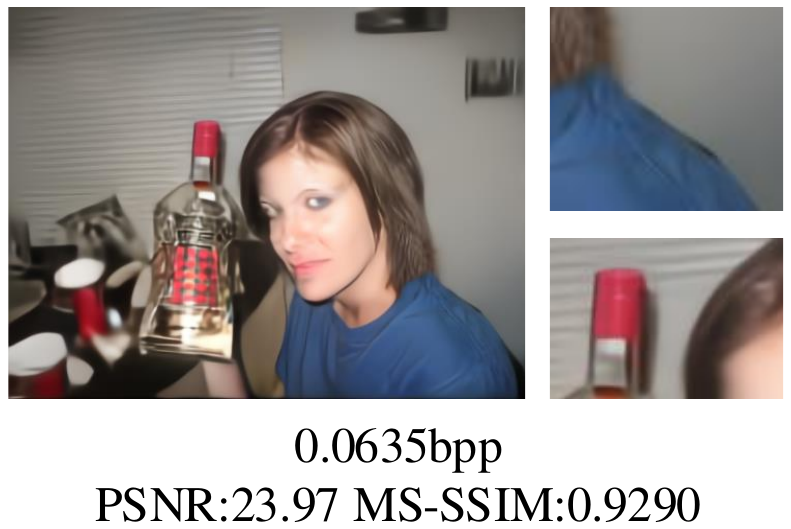}
	\includegraphics[scale=0.5]{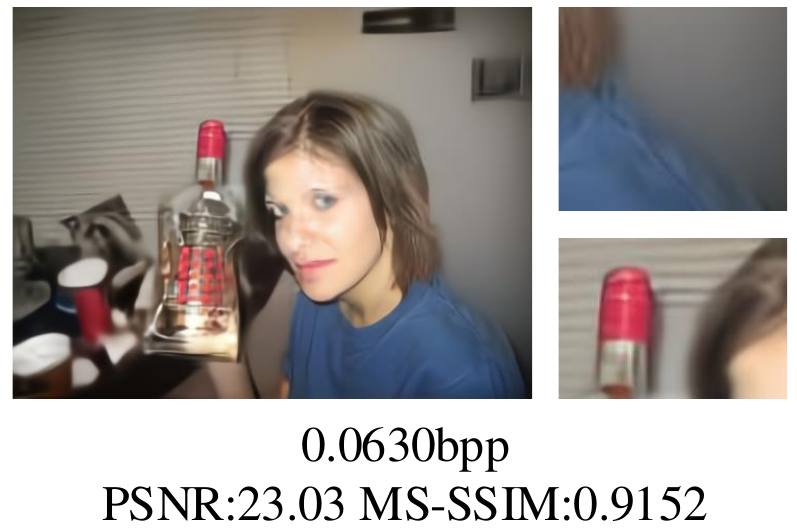}
	\caption{{\bf Masking.} {\it Left}: Masking in feature domain as shown in Fig.~\ref{fig:framework}; {\it Right}: Masking in pixel domain.}
	\label{fig:edge}
\end{figure}

\section{Experimental Discovery} \label{sec:exp}
\subsection{Training}
We choose PASCAL VOC as our dataset, given that it is widely used for object detection and segmentation. 
Our LearntOBIC is trained using the training set of PASCAL VOC 2007 and PASCAL VOC 2012. Input images are set at 320$\times$320$\times$3, while segmentation mask has the same size of $H\times W$ ($H=W=20$) as the latent fMaps. 
 LearntOBIC is tested using the validation set of PASCAL VOC 2012. 

We first use pre-trained DeepLab with ResNet-34 as backbone,  and pre-trained NLAIC to initialize the LearntOBIC model, and then finetune it using the loss function:
\begin{equation}
L = {\lambda}(1-D_{\tt tot}) + {a_1}R_{\tt bkg} + {a_2}R_{\tt obj}.
\label{loss_func} 
\end{equation}
 Here, $D_{\tt tot}$ is defined above. We choose the MS-SSIM as our distortion metric which is reported to have better correlation with human visual perception, especially at low bit rate~\cite{wang2003multiscale}. {$R_{\tt bkg}$} and {$R_{\tt obj}$} are defined in \eqref{eq:bkg_rate} and \eqref{eq:obj_rate}, respectively. 
 
 By adjusting $\lambda$ we achieve rate-distortion trade-off for a variety of bitrates. We use $a_1$ and $a_2$  to adapt the bit consumption for object and background layers. Here, we set $a_1 > a_2$  to shift bits from background to object. We set learning rate at {$10^{-5}$} in the beginning and clip the value to {$5\times10^{-6}$} after 10 epochs. 
We set the total bitrate lower than 0.1 bits per pixel (bpp) to experiment the low bitrate application.

\subsection{Performance Evaluation}
{\bf Objective Efficiency.} Table~\ref{tab:BDrate} has listed the averaged objective results for all test image samples in PASCAL VOC 2012, in terms of bit rate (ave. bpp), MS-SSIM, and PSNR. At such low bit rate, e.g., $<0.07$ bpp (or $>$340x compression ratio), MS-SSIM offers more meaningful quality measurement close to our subjective sensation~\cite{wang2003multiscale}. Learnt image compression methods, e.g., NLAIC, and LearntOBIC, exhibit quite close perceptual index measured by MS-SSIM~\cite{wang2003multiscale}. and both are better than traditional JPEG2K and BPG.

{\bf Subjective Evaluation.}
We further visualize the reconstructed images that are processed using JPEG2K, BPG, NLAIC, and LearntOBIC in Fig.~\ref{fig:visual_result}.
Snapshots in row \#1 to \#3 are from PASCAL VOC 2012. Our LearntOBIC offers significantly better visual results than others (and even slightly smaller bit rate), by intelligently distributing bits between object and background layers. For such low bitrate, traditional JPEG2K and BPG have lost the capacity for fine reconstruction, where severe artifacts are presented in column one and column two, impairing the subjective perception clearly. Though default NLAIC outperforms JPEG2K and BPG subjectively, noticeable artifacts (e.g., over-smoothed texture, and color distortion) are still perceivable. 

\begin{figure}[t]
\centering
	\includegraphics[scale=0.6]{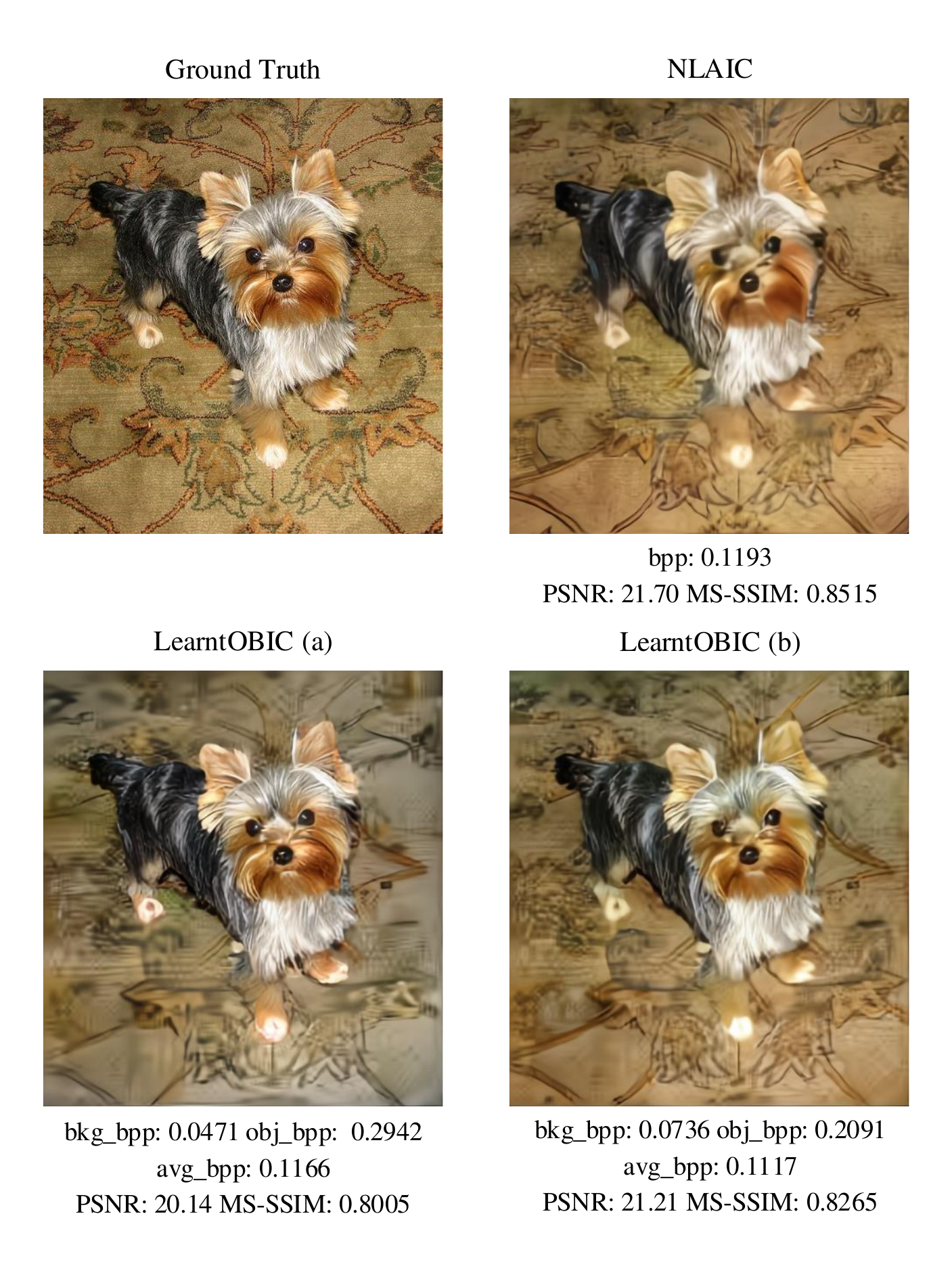}
	\caption{{\bf Visual Comparison for Bit Allocation.} Illustration of quality impact when shifting bits from background to object. {\bf Zoom in the pictures, and you will see more image details.}}
	\label{fig:bit_distribution}
\end{figure}
\subsection{Ablation Studies}
{\bf Masking.} Note that we can also perform the masking in pixel domain to separate object and background layers prior to being fed into compression networks. However, pixel domain masking provides more blurry object boundary, as shown in Fig.~\ref{fig:edge} due to convolutions in compression network applied for deactivated  (e.g., 0 after masking) and activated pixels across the object boundary. Boundary extension or padding may resolve this issue to some extent which is for  our future study.

{\bf Bits Allocation.} We further explore the relationship between bits distribution and subjective quality in LearntOBIC, as shown in Fig.~\ref{fig:bit_distribution}. Given the same total bit rate, by weighing more attention to foreground object with more bits, e.g., from Fig.~\ref{fig:bit_distribution}(b) to Fig.~\ref{fig:bit_distribution}(a), its texture can be reconstructed with finer details, but background is deteriorated with slightly color distortion. This would generally be of interest for applications with the focus on the object, rather the entire image. An interesting exploration avenue is how to automatically shift bits for task oriented applications.

{\bf Model Generalization.} As in Fig.~\ref{fig:visual_result}, e.g., row\#4, \#5, we extend PASCAL trained model to Kodak dataset directly. Our LearntOBIC still provides better visual reconstruction for Kodak images with clearly distinguishable object. As revealed in our work, a more robust segmentation method is highly desired for reliable performance.

\section{Conclusion} \label{sec:concluding_remarks}
We proposed an object-based image compression method, referred to as the LearntOBIC, by integrating the segmentation network and compression network in an end-to-end learnable framework. With this learning-based approach, we could offer the element-wise operations (e.g., masking, convolution-based transforms, etc) to efficiently support the processing of arbitrary-shaped objects. 
Compared with traditional JPEG2K, HEVC-based BPG, as well as the recent learning-based NLAIC, our LearntOBIC offers much improved visual quality for the application scenarios at a very low bitrate.

The OBIC itself is an interesting problem since both human beings and machines are weighing more attentions on particular/salient objects within an image, rather the entire scene, for task oriented application.
This work is our preliminary attempt to revisit the classic OBIC defined almost two decades ago. There are a lot of interesting problems for further investigation. For example, segmentation is generally content dependent. How to make it more robust in this LearntOBIC framework is worth for deep study. On the other hand, how to use sub-stream that corresponds to the image object layer, and how to distribute bits intelligently, are crucial for object-based visual tasks.

\begin{small}
\bibliographystyle{IEEEtran}
\bibliography{object_coding_v2}

\begin{thebibliography}{10}
\providecommand{\url}[1]{#1}
\csname url@samestyle\endcsname
\providecommand{\newblock}{\relax}
\providecommand{\bibinfo}[2]{#2}
\providecommand{\BIBentrySTDinterwordspacing}{\spaceskip=0pt\relax}
\providecommand{\BIBentryALTinterwordstretchfactor}{4}
\providecommand{\BIBentryALTinterwordspacing}{\spaceskip=\fontdimen2\font plus
\BIBentryALTinterwordstretchfactor\fontdimen3\font minus
  \fontdimen4\font\relax}
\providecommand{\BIBforeignlanguage}[2]{{%
\expandafter\ifx\csname l@#1\endcsname\relax
\typeout{** WARNING: IEEEtran.bst: No hyphenation pattern has been}%
\typeout{** loaded for the language `#1'. Using the pattern for}%
\typeout{** the default language instead.}%
\else
\language=\csname l@#1\endcsname
\fi
#2}}
\providecommand{\BIBdecl}{\relax}
\BIBdecl

\bibitem{puri1998mpeg}
A.~Puri and A.~Eleftheriadis, ``{MPEG-4}: An object-based multimedia coding
  standard supporting mobile applications,'' \emph{Mobile Networks and
  Applications}, vol.~3, no.~1, pp. 5--32, 1998.

\bibitem{sikora1995shape}
T.~Sikora and B.~Makai, ``Shape-adaptive {DCT} for generic coding of video,''
  \emph{IEEE Trans. Circuits and Systems for Video Technol.}, vol.~5, no.~1,
  pp. 59--62, 1995.

\bibitem{li2000shape}
S.~Li and W.~Li, ``Shape-adaptive discrete wavelet transforms for arbitrarily
  shaped visual object coding,'' \emph{IEEE Trans. Circuits and Systems for
  Video Technol.}, vol.~10, no.~5, pp. 725--743, 2000.

\bibitem{lecun2015deep}
Y.~LeCun, Y.~Bengio, and G.~Hinton, ``Deep learning,'' \emph{nature}, vol. 521,
  no. 7553, pp. 436--444, 2015.

\bibitem{chen2017deeplab}
L.-C. Chen, G.~Papandreou, I.~Kokkinos, K.~Murphy, and A.~L. Yuille, ``Deeplab:
  Semantic image segmentation with deep convolutional nets, atrous convolution,
  and fully connected crfs,'' \emph{IEEE Trans. pattern analysis and machine
  intelligence}, vol.~40, no.~4, pp. 834--848, 2017.

\bibitem{he2016deep}
K.~He, X.~Zhang, S.~Ren, and J.~Sun, ``Deep residual learning for image
  recognition,'' in \emph{Proceedings of the IEEE CVPR}, 2016, pp. 770--778.

\bibitem{chen2019neural}
T.~Chen, H.~Liu, Z.~Ma, Q.~Shen, X.~Cao, and Y.~Wang, ``Neural image
  compression via non-local attention optimization and improved context
  modeling,'' \emph{arXiv preprint arXiv:1910.06244}, 2019.

\bibitem{skodras2001jpeg}
A.~Skodras, C.~Christopoulos, and T.~Ebrahimi, ``The jpeg 2000 still image
  compression standard,'' \emph{IEEE Signal processing magazine}, vol.~18,
  no.~5, pp. 36--58, 2001.

\bibitem{sullivan2012overview}
G.~J. Sullivan, J.-R. Ohm, W.-J. Han, and T.~Wiegand, ``Overview of the high
  efficiency video coding {(HEVC)} standard,'' \emph{IEEE Trans. Circuits and
  Systems for video technol.}, vol.~22, no.~12, pp. 1649--1668, 2012.

\bibitem{Long_2015_CVPR}
J.~Long, E.~Shelhamer, and T.~Darrell, ``Fully convolutional networks for
  semantic segmentation,'' in \emph{The IEEE Conference on Computer Vision and
  Pattern Recognition (CVPR)}, June 2015.

\bibitem{ronneberger2015u}
O.~Ronneberger, P.~Fischer, and T.~Brox, ``U-net: Convolutional networks for
  biomedical image segmentation,'' in \emph{MICCAI}.\hskip 1em plus 0.5em minus
  0.4em\relax Springer, 2015, pp. 234--241.

\bibitem{lin2017feature}
T.-Y. Lin, P.~Doll{\'a}r, R.~Girshick, K.~He, B.~Hariharan, and S.~Belongie,
  ``Feature pyramid networks for object detection,'' in \emph{Proceedings of
  the IEEE CVPR}, 2017, pp. 2117--2125.

\bibitem{minnen2018joint}
D.~Minnen, J.~Ball{\'e}, and G.~D. Toderici, ``Joint autoregressive and
  hierarchical priors for learned image compression,'' in \emph{NIPS}, 2018,
  pp. 10\,771--10\,780.

\bibitem{balle2016end}
J.~Ball{\'e}, V.~Laparra, and E.~P. Simoncelli, ``End-to-end optimized image
  compression,'' \emph{arXiv:1611.01704}, 2016.

\bibitem{balle2018variational}
J.~Ball{\'e}, D.~Minnen, S.~Singh, S.~J. Hwang, and N.~Johnston, ``Variational
  image compression with a scale hyperprior,'' \emph{arXiv:1802.01436}, 2018.

\bibitem{li2018learning}
M.~Li, W.~Zuo, S.~Gu, D.~Zhao, and D.~Zhang, ``Learning convolutional networks
  for content-weighted image compression,'' in \emph{Proceedings of the IEEE
  CVPR}, 2018, pp. 3214--3223.

\bibitem{agustsson2019generative}
E.~Agustsson, M.~Tschannen, F.~Mentzer, R.~Timofte, and L.~V. Gool,
  ``Generative adversarial networks for extreme learned image compression,'' in
  \emph{ICCV}, 2019, pp. 221--231.

\bibitem{han}
S.~Han and N.~Vasconcelos, ``Image compression using object-based regions of
  interest,'' in \emph{IEEE ICIP}, 2006, pp. 3097--3100.

\bibitem{han2008object}
------, ``Object-based regions of interest for image compression,'' in
  \emph{IEEE DCC}, 2008, pp. 132--141.

\bibitem{chen2006dynamic}
Z.~Chen, J.~Han, and K.~N. Ngan, ``Dynamic bit allocation for multiple video
  object coding,'' \emph{IEEE Tran. Multimedia}, vol.~8, no.~6, pp. 1117--1124,
  2006.

\bibitem{wang2003multiscale}
Z.~Wang, E.~P. Simoncelli, and A.~C. Bovik, ``Multiscale structural similarity
  for image quality assessment,'' in \emph{The 37-th Asilomar Conf. Signals,
  Systems \& Computers, 2003}, 2003, pp. 1398--1402.

\end{thebibliography}
\end{small}

\end{document}